\documentclass[10pt]{amsart}

\usepackage[a4paper]{anysize}
\usepackage{amssymb}
\usepackage{amsmath}
\usepackage{amsfonts,a4wide}

\newcommand{\sse}{\Leftrightarrow}
\newcommand{\se}{\Rightarrow}
\newcommand{\OO}{{\bf 0}}
\newcommand{\HH}{{\bf H}}
\newcommand{\GG}{{\bf G}}
\newcommand{\uu}{{\bf u}}
\newcommand{\bbR}{\mathbb{R}}
\newcommand{\tr}{\operatorname{tr}}
\newcommand{\cof}{\operatorname{cof}}
\newcommand{\diag}{\operatorname{diag}}
\newcommand{\dive}{\operatorname{div}}
\newcommand{\curl}{\operatorname{curl}}
\newcommand{\grad}{\operatorname{grad}}
\newcommand{\rank}{\operatorname{rank}}

\newtheorem{Teo}{Theorem}[section]
\newtheorem{Prop}[Teo]{Proposition}
\newtheorem{Cor}[Teo]{Corolary}

\begin{document}
\begin{abstract}
We show how to use the quasi-Maxwell formalism to obtain solutions of Einstein's field 
equations corresponding to homogeneous cosmologies -- namely Einstein's 
universe, G\"odel's universe and the Ozsvath-Farnsworth-Kerr class I solutions -- written in frames
for which the associated observers are stationary.
\end{abstract}
%
%
\title{Homogeneous cosmologies from the quasi-Maxwell formalism}
\author{Jo\~ao Costa}
\address{Departmento de M\'etodos Quantitativos, Instituto Superior das Ci\^encias do Trabalho e da Empresa, Portugal}
\author{Jos\'{e} Nat\'{a}rio}
\address{Department of Mathematics, Instituto Superior T\'{e}cnico, Portugal}
\thanks{The second author was partially supported by FCT/POCTI/FEDER}
\maketitle
%
%
%
\section*{Introduction}
A particularly intuitive framework for obtaining and interpreting stationary solutions of 
Einstein's field equations is the so-called {\em quasi-Maxwell formalism} (\cite{LN98}, \cite{O02}).
Although such solutions have been extensively treated in the past (\cite{B00}, \cite{SKMHH03}), 
this approach has been successfully used in recent times (\cite{NZ97}, \cite{NZT01}). 
In this paper we apply the quasi-Maxwell formalism in the case when the 
space manifold is a Lie group with left-invariant metric and fields, and 
rediscover Einstein's universe, G\"odel's universe and the Ozsvath-Farnsworth-Kerr 
class I solutions, sometimes written in unconventional frames.

The organization of the paper is as follows: In the first section we briefly review 
the quasi-Maxwell formalism for stationary spacetimes. In the second section we analyze
the form taken by the quasi-Maxwell equations when the space manifold is a Lie group. In 
the third section we further specialize to Lie groups with class A Lie algebras. Finally, 
solutions of the quasi-Maxwell equations for these space manifold are obtained and identified 
in the last section.

We use Einstein summation convention, irrespective of the position of the indices (which will 
often be irrelevant as we will be leading with orthonormal frames on Riemannian manifolds). We will 
take Latin indices $i,j,\ldots$ to run from $1$ to $3$.
%
%
%
\section{Quasi-Maxwell formalism}
In this section we briefly review the quasi-Maxwell formalism for stationary spacetimes. 
For more details, see \cite{O02}.

Recall that a {\em stationary spacetime} $(M,g)$ is a Lorentzian $4$-manifold with
a global timelike Killing vector field $T$. We assume that there exists a global
time function $t:M\rightarrow\bbR$ such that $T= \frac{\partial}{\partial t}$.
The quotient of $M$ by the integral curves of $T$ is a $3$-dimensional manifold 
$\Sigma$ to which we refer as the {\em space manifold}. If 
$\{x^i\}$ are local coordinates in $\Sigma$, we can write 
the line element of $(M,g)$ as
\[
ds^2= -e^{2\phi}\left(dt+A_idx^i\right)^2+\gamma_{ij}dx^idx^j
\]
where $\phi$, $A_i$ and $\gamma_{ij}$ do not depend on $t$.
This allows us to interpret $\phi$, $A=A_idx^i$ and 
$\gamma=\gamma_{ij}dx^i\otimes dx^j$ as tensor fields on the 
space manifold. It turns out that $\gamma$ is a Riemannian metric in $\Sigma$, 
independent of the choice of the global time function $t$. The differential forms 
$G=-d \phi$ and  $H=-e^ \phi dA$ are also independent of this choice, and play a central role 
in the so-called {\em quasi-Maxwell formalism}. We define 
the {\em gravitational} and {\em gravitomagnetic} 
(vector) fields $\GG$ and $\HH$ through
\begin{align}
& G=\gamma(\GG,\cdot) \\
& H=\epsilon(\HH,\cdot,\cdot)
\end{align}
where $\epsilon$ is a Riemannian volume form in $(\Sigma, \gamma)$ (which we assume to be orientable).

We identify a vector ${\bf v} \in T_p\Sigma$ with the unique vector field $v$ along the integral curve 
of $T$ through $p$ which is orthogonal to $T$ and satisfies $\pi_* v = {\bf v}$ ($\pi:M\to\Sigma$
being the quotient map). Let $\{X_0,X_i \}$ be a local orthonormal frame on $M$, where $X_0 = \left(-g(T,T)\right)^{-\frac12} T$.
If
\[
u=u^0 X_0+u^iX_i = u^0 X_0+\uu
\]
represents the unit tangent vector to a timelike geodesic, the
motion equation
\[
\widetilde{\nabla}_{u}\,u=0
\]
is equivalent to
\[
\nabla_{\uu}\,\uu= u^0\left( u^0\,\GG+\uu\times\HH \right)
\]
\noindent
with $u^0 = \left( 1 + \uu\,^2\right)^\frac12$ (where $\widetilde{\nabla}$ is the Levi-Civita 
connection of $(M,g)$, $\nabla$ is the Levi-Civita connection of $(\Sigma, \gamma)$ and
$\uu\,^2 = \gamma(\uu,\uu)$).

If we let $R_{ij}$ and $\nabla_iG_j$ represent the components of the Ricci tensor of $\nabla$ and of 
the covariant derivative of $G$, Einstein's equations for a perfect fluid with density $\rho$,  
pressure $p$ and $4$-velocity $u$ reduce to the {\em quasi-Maxwell equations} ($QM$)
\begin{align*}
& \dive\GG= \GG\,^2+\frac12\HH\,^{2}-8\pi(\rho+p)\uu\,^2-4\pi(\rho+3p) \tag{$QM.1$} \\
& \curl\HH=2\GG\times\HH-16\pi(\rho+p)u^0\uu \tag{$QM.2$} \\
& R_{ij}+\nabla_iG_j=G_iG_j+\frac12 H_iH_j-\frac12\HH\,^2\gamma_{ij} \tag{$QM.3.ij$} \\
& \hspace{1.9cm} + 8\pi\left((\rho+p)u_iu_j+\frac12(\rho-p)\gamma_{ij}\right). 
\end{align*}

We can use $QM$ to solve Einstein's equations by writing down a 
Riemannian metric for the space manifold (eventually depending on unknown functions), 
and solving for the fields (see \cite{LN98}, \cite{NZ97}, \cite{NZT01}, \cite{O02}).
For instance, the Schwarzchild solution is the 
static solution (i.e., $\HH=0$) obtained when we consider a spherically 
symmetric space manifold with radial $\GG$.

A word of caution must be issued here: the quasi-Maxwell decomposition does depend on 
the choice of the timelike Killing vector field $T$. Therefore when one solves the $QM$ 
equations, one is really solving for $(M,g,T)$. If a given spacetime has a large enough 
isometry group, it can yield many different solutions of $QM$.

The goal of this paper is the classification of solutions whose space manifolds are  
Lie groups with left-invariant metrics, and whose vector fields $\GG$ and $\HH$ are
left-invariant.
%
%
%
\section{Quasi-Maxwell equations for a Lie group}
Let the space manifold $\Sigma$ be a $3$-dimensional Lie group. To choose 
a left-invariant metric we fix a frame $\{X_i\}$ of left-invariant vector fields 
and declare it to be orthonormal. All the information about the geometry of 
the space manifold will then be encoded in the structure constants, defined by
\[
\left[X_i,X_j\right] = C^k_{ij}X_k = C_{kij}X_k.
\]
The last equality emphasizes that there is no need to worry about the vertical position 
of the indices, as we're working with an orthonormal frame. The Christoffel symbols of the 
Levi-Civita connection are then given by
\[
\Gamma^i_{jk}=\frac12\left(C_{ijk}+C_{kij}-C_{jki}\right).
\]
Letting $G=G_i\omega^i$, where $\{\omega^i\}$ is the dual basis of $\{X_i\}$, we have
\[
\nabla_iG_j  = - \Gamma^k_{ij}\,G_k.
\]
Consequently,
\[
\dive\GG=\nabla_iG_i=-\Gamma^k_{ii}G_k.
\]
The Maurer-Cartan formula
\[
d\omega^i=-{\frac12}C^i_{jk}\omega^j\wedge\omega^k
\]
assures us that the exterior derivative is a linear transformation between the spaces $\Omega^1_L(\Sigma)$ and 
$\Omega^2_L(\Sigma)$ of the left-invariant $1$ and $2$-forms, whose 
matrix for the bases $\{\omega^1,\omega^2,\omega^3\}$ of $\Omega^1_L(\Sigma)$ 
and $\{\omega^2\wedge\omega^3,\omega^3\wedge\omega^1,\omega^1\wedge\omega^2\}$ of $\Omega^2_L(\Sigma)$ is
\[
D = \left(
\begin{array}{ccc}
C_{132} & C_{232}& C_{332}\\
C_{113} & C_{213}& C_{313}\\
C_{121} & C_{221}& C_{321}
\end{array}\right).
\]
By definition, curl$\,\HH$ is the only vector field satisfying
\[
\epsilon(\curl\HH,\cdot,\cdot)=d(\gamma(\HH,\cdot)).
\]
\noindent
Since vectors, $1$ and $2$-forms are related by the
isomorphisms given by the metric and the volume element of $\Sigma$, we obtain
\[
\curl\HH= \left(X _1 \,\,\,\, X _2 \,\,\,\, X _3 \right)
\cdot  D \cdot
\left(
\begin{array}{ccc}
H_1\\
H_2\\
H_3
\end{array}\right).
\]

The fact that $\gamma$, $\GG$ and $\HH$ are left-invariant imposes restrictions 
on the fluid generating the gravitational field:

\begin{Prop} 
The density and pressure are constant and $\uu$ is left-invariant.
\end{Prop}

\begin{proof}
$QM.3.ii$ gives us
\[
(\rho+p)u_i^2=-{\frac12}(\rho-p)+\text{constant}.
\]
Adding these three equations we obtain
\[
(\rho+p)\uu^{\,2}=-{\frac32}(\rho-p)+\text{constant},
\]
which substituted in $QM.1$ yields
\[
-3(\rho-p) + \rho + 3p=\text{constant} \sse 3p-\rho = \text{constant}.
\]
From $QM.2$ we see that
\begin{align*}
(\rho+p)u^0u_i= \text{constant} & \se (\rho+p)^2 \left(u^0\right)^2u_i^{\,2}= \text{constant} \\
& \sse (\rho+p) \left(\uu^{\,2}+1\right)(\rho+p)u_i^{\,2} = \text{constant}\\
& \sse \left[-{\frac32}(\rho-p)+\text{const.}+(\rho+p)\right] \cdot \left[-{\frac12}(\rho-p)+\text{const.}\right]=\text{const.} \\
& \sse {\frac14}\rho^{\,2}+{\frac54}p^{\,2}-{\frac32}\rho p + \text{first order terms} = \text{constant}.
\end{align*}
But since $\rho=3p+ \text{constant}$, we get
\[
-p^2+\text{first order terms} = \text{constant}.
\]
We conclude that $\rho$ and $p$ can take at most two distinct values in $\Sigma$, and, being so, the result follows from their continuity.

Its now clear that for $\rho+p\neq0$ the components of $\uu$ are constant, which suffices to insure that
it is left-invariant. For $\rho+p=0$, $\uu$ becomes undefined and we can take it to be left-invariant (e.g. zero) without
loss of generality.
\end{proof}

\begin{Cor} The  vector field $u$ has the following proprieties:
\[
\widetilde{\dive}\,\,u=0
\]
and
\[
\widetilde{\nabla}_{u}\,u=0.
\]
\end{Cor}

\begin{proof}
We have seen that we only have to consider the case $\rho+p\neq0$. Euler's equation for a perfect fluid is
\[
\widetilde{\dive}\,T=0 \sse
\left\{
\begin{array}{l}
\widetilde{\dive}\left(\rho\,u\right)+p\,\widetilde{\dive}\,u=0 \\
(\rho+p)\widetilde{\nabla}_{u}\,u=-\left(\widetilde{\grad}\,p\right)^{\perp}
\end{array}
\right.
\]
where $\left(\widetilde{\grad}\,p\right)^{\perp}$ designates the component of $\widetilde{\grad}\,p$ 
orthogonal to $u$. Since $\rho$ and $p$ are constant with $\rho+p\neq0$, it follows that
\[
\left\{
\begin{array}{l}
(\rho+p)\,\widetilde{\dive}\,u=0 \\
(\rho+p)\,\widetilde{\nabla}_{u}\,u=0
\end{array}
\right. 
\sse 
\left\{
\begin{array}{l}
\widetilde{\dive}\,u=0 \\
\widetilde{\nabla}_{u}\,u=0
\end{array}
\right..
\]
\end{proof}

\begin{Cor} 
The vector fields $\uu$ and $\GG$ are orthogonal.
\end{Cor}

\begin{proof}
The motion equation yields
\[
\widetilde{\nabla}_{u}\,u=0 \se
\nabla_{\uu}\,\uu= u^0\left( u^0\,\GG+\uu\times\HH \right).
\]
Since $\uu$ is left-invariant and $u^0$ is a nonzero constant,
\[
0 = \frac{d}{d\tau} \gamma(\uu,\uu) = 2 \gamma(\nabla_{\uu} \uu, \uu) = 2(u^0)^2 \gamma(\GG,\uu)
\quad \se \quad \gamma(\GG,\uu) = 0.
\]
\end{proof}

The following result relates solutions corresponding to conformally related 
left-invariant metrics:

\begin{Prop} {\bf (Rescaling Lemma)} \label{rescaling}
From a solution $(G_i,H_j,u_k,\rho,p)$ of $QM$, where the left-invariant metric is associated to the frame 
$\{X_i\}$, we can construct a solution $(\widehat{G}_i,\widehat{H}_j,\widehat{u}_k,\widehat{\rho},\widehat{p}\,)$ 
for the left-invariant metric associated to the frame $\{\widehat{X}_i=\lambda X_i\}$, by setting
\begin{align*}
& \widehat{G}_i=\lambda G_i \\
& \widehat{H}_j=\lambda H_j \\
& \widehat{u}_k = u_k \\
& \widehat{\rho} =\lambda^2\rho \\ & \widehat{p} =\lambda^2 p.
\end{align*}
\end{Prop}

\begin{proof}
Since
\[
[\widehat{X}_i,\widehat{X}_j]=\lambda^2[X_i,X_j]=\lambda^2 C_{kij}X_k = \lambda C_{kij}\widehat{X}_k,
\]
we obtain 
\[
\widehat{C}_{kij}=\lambda C_{kij},
\]
from which follows
\[
\widehat{\Gamma}^i_{jk}=\lambda \Gamma^i_{jk} \quad \text{and} \quad \widehat{D}= \lambda D
\]
and consequently
\[
\widehat{R}_{ij}=\lambda^2 R_{ij}, \quad
\widehat{\dive}\,\widehat{G}=\lambda^2 \dive\,G \quad
\text{and} \quad
\widehat{\curl}\,\widehat{H}=\lambda^2 \curl\,H.
\]
Since that, by construction, $\widehat{\gamma}_{ij}=\gamma_{ij}= \delta_{ij}$, 
the result follows.
\end{proof}

It is easy to see that this rescaling corresponds to rescaling the full 
spacetime metric by $\frac1{\lambda^2}$.
%
%
%
\section{Class A Lie Algebras}
If we take $\Sigma$ to be connected and simply connected, Lie's theorem \cite{W83} guarantees 
that the space manifold will be uniquely determined, up to isomorphism, by its Lie algebra. 
Therefore, the consideration of all possible space manifolds becomes the classification 
of three-dimensional Lie algebras - a much simpler task!

Following \cite{W84}, we learn that this classification may be realized 
by means of a $2\choose0$ symmetric tensor $M$ and a covector
$\nu\in\ker M$, whose components in a given basis for 
the Lie algebra are $\nu_i=C^k_{ki}$. It becomes natural to divide the 
classification in two classes: class A for Lie algebras with 
$\nu=0$, and class $B$ for Lie algebras with $\nu\neq0$.

We shall restrict ourselves to class A algebras. These are classified by 
the rank and signature of the symmetric tensor $M$, and are six in total: the 
abelian algebra (corresponding to $\rank M = 0$), the 
Heisenberg algebra (corresponding to $\rank M = 1$), the semidirect products 
$\mathfrak{so}(1,1) \ltimes \bbR^2$ and $\mathfrak{so}(2) \ltimes \bbR^2$
(corresponding to the two possible signatures for $\rank M = 2$) and the 
simple algebras $\mathfrak{sl}(2)$ and $\mathfrak{so}(3)$ (corresponding to 
the two possible signatures for $\rank M = 3$). In terms of the more usual 
Bianchi classification, these are Bianchi types I, II, VI with parameter $h=-1$, 
VII with parameter $h=0$, VIII and IX, respectively (see \cite{RS75}, \cite{SKMHH03}).

Since $C^k_{ki}=0$, $M$ can be identified, using the left-invariant metric
on which the Lie algebra basis is orthonormal, with minus the linear operator 
yielding the exterior derivative restricted to $\Omega^1_L$.
Therefore, class A Lie algebras are classified by the rank and 
signature of the matrix $D$ of the previous section. This matrix is also
useful for computing the Ricci tensor:

\begin{Prop} \label{Ricci}
In a Lie group with class A Lie algebra and left-invariant 
metric, the matrix of components of the Ricci tensor in 
the basis $\{\omega^i\otimes\omega^j\}$, where $\{\omega^i\}$ is an 
orthonormal left-invariant co-frame, is given by
\[
\left( R_{ij} \right) =  D^2-{\frac12}\tr\left(D^2\right)I+\cof\left(D\right).
\]
\end{Prop}

The proof of this result is straightforward but lengthy and
will be omitted.

Since $D$ is symmetric, we are guaranteed the existence of 
a left-invariant orthonormal co-frame $\{\omega_i\}$ for which
\[
D=\diag(C_{132},C_{213},C_{321}).
\]
Consequently, we can eliminate two unknowns in $QM$:

\begin{Prop} \label{G=GX1}
There exists a left-invariant orthonormal frame $\{\widehat{X}_i\}$ for which 
the exterior derivative matrix in the basis  $\{\gamma(\widehat{X}_k,\cdot)\}$ 
and $\{\epsilon(\widehat{X}_k,\cdot,\cdot)\}$ is diagonal and $\GG=G\widehat{X}_1$.
\end{Prop}

\begin{proof}
Choose  $\{X_i\}$ such that $D=\diag(a,b,c)$ and let $\GG=G_i X_i$. Since $G$ 
is a closed 1-form, we get
\[
dG=d(\gamma(\GG,\cdot))=0 \sse aG_1X_1+bG_2X_2+cG_3X_3=0.
\]

Rearranging  the indices if necessary, the last equation tells us that:

\begin{enumerate}
\item
$\rank(D) = 3 \se a,b,c\neq0 \se \GG=\OO$;
\item
$\rank(D)=2 \se a=0,\,b,c\neq0 \se G_2=G_3=0$;
\item \label{r1}
$\rank(D)=1 \se a,b=0,\,c\neq0 \se G_3=0 \se \GG\perp X_3$. For the nontrivial case (i.e., $\GG\neq\OO$) 
it suffices to choose  $\widehat{X}_1 = \frac{1}{\|\GG\|}\GG$, $\widehat{X}_3=X_3$ and $\widehat{X}_2$ in such a way as
 to complete the basis as an orthonormal basis;
\item
$\rank(D)=0$: identical to (\ref{r1}).
\end{enumerate}
\end{proof}

We end this section with three useful results easily proved from 
the diagonalization of the exterior derivative matrix.

\begin{Prop} 
Left-invariant vector fields have vanishing divergence.
\end{Prop}

\begin{proof}
If we choose a basis for which $D$ is diagonal,
we conclude that the only structure constants not necessarily 
zero are those with no repeated indices, and consequently
\[
\Gamma^i_{jk} \neq 0 \se (i,j,k) \text{ is a permutation of } (1,2,3).
\]
The result then follows from the equation
\[
\dive\,\GG=-\Gamma^k_{ii}G_k.
\]
\end{proof}

Equivalently, we have

\begin{Prop} 
$d(\Omega^2_L)=0$.
\end{Prop}

\begin{Cor} \label{orthog}
$\GG$ and $\HH$ are orthogonal.
\end{Cor}

\begin{proof}
Since $H$ is a left-invariant 2-form, the last result tells us that
\begin{align*}
dH=0 & \sse d \left(-e^{\phi}dA \right)=0 \\
& \sse -e^{\phi}d\phi \wedge dA-e^{\phi}d(dA)=0 \\
& \sse G \wedge H =0.
\end{align*}
Using proposition \ref{G=GX1}, we get
\begin{align*}
& G_1 \omega^1 \wedge (H_1 \omega^2 \wedge \omega^3 +H_2 \omega^3 \wedge \omega^1+H_3 \omega^1 \wedge \omega^2)=0 \\
& \sse G_1H_1 \, \omega^1 \wedge \omega^2\wedge\omega^3=0 \sse G_1H_1=0 \sse \gamma(\GG,\HH)=0.
\end{align*}
\end{proof}
%
%
%
\section {Classification}
For now on we will consider only orthonormal bases $\{X_i\}$ of left-invariant 
vector fields for the class $A$ Lie algebras of the space 
manifold such that  $D=\diag(a,b,c)$. From Proposition \ref{Ricci}
we have
\[
(R_{ij})=\diag\left(\frac12 a^2-\frac12 b^2-\frac12 c^2 + bc\,,
-\frac12 a^2+\frac12 b^2-\frac12 c^2+ac\,,-\frac12 a^2-\frac12 b^2+\frac12 c^2 +ab\right).
\]
%
%
%
\subsection{Vacuum solutions with cosmological constant.}
For convenience, we begin with the computation of $QM$
solutions such that $\rho+p=0$. These correspond to vacuum 
solutions with cosmological constant.

\begin{Prop} \label{cconst}
The only  $QM$ vacuum solution with cosmological 
constant ($\rho + p =0$) is Minkowski spacetime, i.e.,
$\GG=\HH=0$ and $\rho=p=0$. 
The space manifold is then Ricci-flat ($Ricci=0$), and hence we 
necessarily have $D=\diag(0,b,b)$ for some $b\in \bbR$
in an appropriate basis of the space manifold's Lie algebra.
\end{Prop}

\begin{proof}
Let $\rho + p =0$. The indefiniteness of
$\uu$ allows us to assume, without loss of generality (wlg),
that $\uu=0$. From the motion equation we get
\[
0=\left(u^0 \right)^2\GG=0 \sse \GG=0.
\]
Therefore,
\[
QM.1 \sse 0 = \frac12 \HH^2 - 4\pi(\rho+3p) \sse \HH^2 = 16\pi p.
\]
Since $(R_{ij})$ is diagonal,
\[
QM.3.ij \,\, (i\neq j) \sse 0 = H_iH_j.
\]
Therefore, two of the components of $\HH$ must vanish. Taking,
wlg, $\HH=HX_1$ and writing $D=\diag(a,b,c)$, we get
\[
QM.2 \sse D \cdot \HH =0 \sse aH=0.
\]

If $H=0$, we obtain $p=0 \se \rho=0$.

If $a=0$,
\[
QM.3.ii \,\, (i\neq1) \sse R_{22}=R_{33}=
-\frac12\HH^2+4\pi(\rho-p) = 4\pi(\rho-3p).
\]
But
\[
R_{22}=R_{33} \sse \frac12 b^2 -\frac12 c^2 = -\frac12 b^2 +\frac12
 c^2 \sse b^2=c^2 \se R_{22}=R_{33}=0,
\]
yielding $\rho-3p=0$, and therefore $\rho=p=0$ (hence $H=0$).

Thus the only solution with $\rho+p=0$ is Minkowski spacetime, and 
verifies $Ricci=0$. From the diagonalization of $D$, 
it is easily seen that a space manifold is Ricci-flat if and only if there is 
a basis for its Lie algebra such that $D=\diag(0,b,b)$, $b\in\bbR$.
\end{proof}

For the remaining computations we will therefore assume that $\rho+p\neq0$.
%
%
%
\subsection{Solutions with a flat space manifold.}

In this section we will compute all solutions of $QM$ with flat space manifold $(\Sigma,\gamma)$. 
Since this is a $3$-dimensional manifold, the curvature tensor is completely 
determined by the Ricci tensor, and therefore flatness is equivalent to Ricci-flatness.

\begin{Teo} \label{flat}
The $QM$ solutions with flat space manifold (i.e., with $Ricci=0$) 
correspond to Lie algebras with a basis for which $D=\diag(0,b,b)$, 
$b\in \bbR$, and such that:
\begin{enumerate}
 \item
(G\"odel's universe) $b=0$, $\GG=\sqrt{16\pi p}\,X_1$, $\HH=\sqrt{32\pi p}\,X_2$, 
$\uu=X_3$ and $\rho=p \in \bbR^+$;
\item 
(Minkowski spacetime) $\GG=\HH=\OO$, $p=\rho=0$ is a solution, for all $b\in\bbR$ (cf. Proposition \ref{cconst}).
\end{enumerate}
\end{Teo}

\begin{proof}
We already saw that Ricci-flatness implies that we can choose $D=\diag(0,b,b)$, $b\in \bbR$. 
Arguing as in the demonstration of proposition \ref{G=GX1}, we can take $\GG=GX_1$ and $\HH=H_1X_1+H_2X_2$.

Suppose first that $\GG=\OO$. In this case,
\[
QM.2 \sse 
\left\{
\begin{array}{l}
0=16\pi(\rho+p)u^0u_1 \\
bH_2=-16\pi(\rho+p)u^0u_2 \\
0=16\pi(\rho+p)u^0u_3
\end{array}
\right. 
\se u_1=u_3=0 \sse \uu=uX_2.
\]

We then have as the only non trivial equation $QM.3.ij \,\,(i\neq j)$
\[
QM.3.12 \sse H_1H_2=0.
\]

If  $H_1=0$, then $\HH=HX_2$. Therefore
\begin{align*}
QM.3.ii & \sse
\left\{
\begin{array}{l}
0=-\frac12 H^2 +4\pi(\rho-p) \\
0 = \frac12 H^2-\frac12 H^2+8\pi(\rho+p)u^2+4\pi(\rho-p) \\
0=-\frac12 H^2 +4\pi(\rho-p)
\end{array}
\right. \\
& \sse
\left\{
\begin{array}{l}
H^2=8\pi(\rho-p) \\
8\pi(\rho+p)u^2=-4\pi(\rho-p)=-\frac12 H^2
\end{array}
\right.
\end{align*}
and
\[
QM.1 \sse 0= \frac12 H^2 - 8\pi(\rho+p)u^2-4\pi(\rho+3p) \sse
H^2=4\pi(\rho+3p).
\]
Consequently,
\[
4\pi(\rho+3p)=8\pi(\rho-p) \sse \rho = 5p,
\]
and therefore
\[
QM3.22 \sse 8\pi(\rho+p)u^2=-4\pi(\rho-p) \sse 
12 p u^2 = -4 p \se p=0\se \rho = 0.
\]

If $H_2=0 \left(\se \HH=HX_1\right)$, we get
\begin{align*}
QM.3.ii & \sse
\left\{
\begin{array}{l}
0=4\pi(\rho-p) \\
0 = - \frac12 H^2 +8\pi(\rho+p)u^2 +4\pi(\rho-p) \\
0=- \frac12 H^2  +4\pi(\rho-p)
\end{array}
\right. \\
& \sse 
\left\{
\begin{array}{l}
\rho=p \\
\uu=\HH=\OO
\end{array}
\right. 
\end{align*}
and
\[
QM.1 \sse 0=4\pi(\rho+3p).
\]
But since $\rho=p$, we obtain $\rho=p=0$.

Let us now consider the case $\GG\neq\OO$. From corollary  \ref{orthog} we have
\[
\gamma(\GG,\HH)=0 \sse H_1=0 \sse \HH = HX_2.
\]

If $b=0$,
\[
QM.2 \sse
\left\{
\begin{array}{l}
0=u_1 \\
0=u_2 \\
0 =2GH-16\pi(\rho+p)u^0u_3
\end{array}
\right.
\sse
\left\{
\begin{array}{l}
\uu=uX_3 \\
GH = 8\pi(\rho+p)u^0u
\end{array}
\right.,
\]
and since
\begin{align*}
\nabla G &= \nabla_iG_j\,\omega^i\otimes\omega^j \\
&= -\Gamma^k_{ij}G_k\,\omega^i\otimes\omega^j\\
&= -\Gamma^1_{23}G_1\,\omega^2\otimes\omega^3-\Gamma^1_{32}G_1\,\omega^3\otimes\omega^2\\
&= -\frac12(C_{123}+C_{312}-C_{231})G\,\omega^2\otimes\omega^3-\frac12(C_{132}+C_{213}-C_{321})G\,\omega^3\otimes\omega^2\\
&= -\frac12(0-b+b)G\,\omega^2\otimes\omega^3-\frac12(0+b-b)G\,\omega^3\otimes\omega^2 = 0,
\end{align*}
equations $QM.3.ij \,\,(i\neq j)$ are trivial.

On the other hand,
\begin{align*}
QM.3.ii & \sse
\left\{
\begin{array}{l}
0=G^2 - \frac12 H^2 +4 \pi (\rho-p) \\
0 =4 \pi (\rho-p)\\
0 = - \frac12 H^2 +8 \pi (\rho+p)u^2+4 \pi (\rho-p)
\end{array}
\right. \\
&\sse
\left\{
\begin{array}{l}
G^2=\frac12 H^2 \se H \neq0 \\
\rho=p \\
16 \pi p\,u^2 =\frac12 H^2
\end{array}
\right.
\end{align*}
from which
\begin{align*}
QM.1 & \sse 0 = G^2 + \frac12 H^2 -8 \pi (\rho+p)u^2-4 \pi (\rho+3p)\\
& \sse G^2=4 \pi (\rho+3p)= 16\pi p \se H^2=32\pi p \quad \text{and} \quad p>0.
\end{align*}
Consequently,
\[
16\pi p \, u^2 = \frac12 H^2 = 16\pi p \sse u^2=1.
\]
Equation $QM.2.3$ is immediately satisfied if we respect
its only imposition: $G H u >0$. It can be shown that
this solution is in fact G\"odel's universe (see section \ref{ID}).

We are now left with the case $\GG \neq \OO$, $b \neq 0$. We have
\[
QM.2 \sse
\left\{
\begin{array}{l}
0=u_1 \\
bH = -16 \pi (\rho+p) u^0u_2 \\
0 =2GH-16\pi(\rho+p)u^0u_3
\end{array}
\right.
\]
Since $\nabla G=0$, $\HH=HX_2$ and $u_1=0$, all of the $QM.3.ij\,\,(i\neq j)$ 
are trivial with the exception of
\begin{align*}
QM.3.23 \sse QM.3.32  & \sse 0 =  8\pi(\rho+p)u_2u_3\\ 
&\sse 0 = u_2u_3.
\end{align*}
But since the components of $\uu$ are constant,
\[
\nabla_{\uu}\,\uu=u_iu_j\nabla_{X_i}X_j=u_2u_3\left(\nabla_{X_2}X_3+\nabla_{X_3}X_2\right) =0.
\]

If $u_3=0$ we obtain $\uu$ parallel to $\HH$ and hence
\[
\text{Motion Equation } \sse 0=\left(u^0\right)^2G \sse G=0,
\]
yielding a contradiction.

If $u_2=0$, $QM.2.2 \se H=0$ and again the motion equation will lead us to $G=0$. Therefore 
we must have $b=0$ whenever $\GG \neq \OO$.
\end{proof}
%
%
%
\subsection{Solutions for Lie algebras with $\boldsymbol{\rank D=3}$.}
It is easily seen that a change of basis from $\{ X_1, X_2, X_3 \}$ to $\{ -X_1, X_2, X_3 \}$ 
changes the exterior derivative matrix from $D = \diag(a,b,c)$ to $D = \diag(-a,-b,-c)$. Therefore 
we can assume wlg that $a>0$.

\begin{Teo} \label{rank3}
The $QM$ solutions with $\rank D = 3$ correspond to Lie algebras 
with a basis such that $a>0$ and:
\begin{enumerate}
\item
(Einstein's universe) $D=\diag(a,b,b)$, with $b>0$, 
$a \geq b$, $\GG=\OO$, $\HH=\sqrt{a(a-b)}\,X_1$, $\uu = -\sqrt{\frac{a-b}{b}}\,X_1$ and 
$\rho=-3p=\frac{3ab}{32\pi}$;
\item
(G\"odel's universe) $D=\diag(a,b,b)$, with $b<0$, $\GG=0$, 
$\HH=\sqrt{a(a-2b)}\,X_1$, $\uu = -\sqrt{-\frac{a}{2b}}\,X_1$ and
$\rho=p= -\frac{ab}{16\pi}$;
\item 
(Ozsvath-Farnsworth-Kerr class I) $D=\diag(a,b,a-b)$, with $16b(a-b)>3a^2 \sse \frac14 a < b < \frac34 a$, $\GG=0$, $\HH= \sqrt{4b(a-b)-\frac12 a^2}\,X_1$, 
$\uu= - \frac{a}{\sqrt{16b(a-b)-3a^2}}\,X_1$,
$p=-\frac{a^2}{64\pi}$ and $\rho=\frac{32b(a-b)-5a^2}{64\pi}$.
\end{enumerate}
\end{Teo}

\begin{proof}
Let $D=\diag(d_1,d_2,d_3)$, with $\Pi_id_i \neq0$. Then $\GG=0$, and consequently
\[
QM.2 \sse d_iH_i=-16\pi(\rho +p)u^0u_i \sse H_i=-\frac{16\pi(\rho+p)}{d_i} \,u^0u_i
\]
(the Einstein summation convention will not apply for the duration of this proof). Therefore,

\begin{align*}
QM.3.ij \,\,(i \neq j) & \sse  0=H_iH_j + 16 \pi (\rho+p)u_iu_j \\
&\sse 0 = \frac{\left[16 \pi (\rho+p)\right]^2}{d_id_j} \,\left(u^0 \right)^2u_iu_j+  16 \pi (\rho+p)u_iu_j \\
& \sse 0= 16 \pi(\rho+p)u_iu_j \left(\frac{16 \pi (\rho+p)}{d_id_j}\left(u^0 \right)^2+1 \right ) \\
& \sse 0=u_iu_j \quad \text{or} \quad \left(u^0 \right)^2 = -\frac{d_id_j}{16\pi(\rho +p)}.
\end{align*}

We have to consider the following cases:

\begin{enumerate}
\item
$u_{i_1}=0$ and:
\begin{enumerate}
\item
$u_{i_2}=0$;
\item
$\left(u^0 \right)^2 = -\frac{d_{i_2}d_{i_3}}{16\pi(\rho +p)}$
\end{enumerate}
(where $(i_1, i_2, i_3)$ is an arbitrary permutation of $(1,2,3)$);
\item
$\left(u^0 \right)^2 = -\frac{d_id_j}{16\pi(\rho +p)}$, for all $i,j \in \{1,2,3\}$ with $i \neq j$.
\end{enumerate}

Let us do so:

\begin{enumerate}
\item
\begin{enumerate}
\item
Suppose, wlg, that $u_2=u_3=0 \se \uu=uX_1$. Then $QM.2 \se \HH=HX_1$, and therefore
\[
QM.3.ii \,\, (i \neq1) \sse R_{22}=R_{33}=-\frac12 H^2+4\pi(\rho-p).
\]
However,
\begin{align*}
R_{22}=R_{33} & \sse -\frac12 a^2 + \frac12 b^2 -\frac12 c^2 +ac = -\frac12 a^2 - \frac12 b^2 +\frac12 c^2 +ab  \\
& \sse b^2-c^2+ac-ab =0\\
& \sse (b-c)(b+c)-a(b-c)=0 \\
& \sse (b-c)(b+c-a)=0 \\
& \sse c=b \quad \text{or} \quad c=a-b,
\end{align*}
which  leads us to the consideration of two sub-cases:
\begin{enumerate}
\item
$c=b$;
\item
$c=a-b$.
\end{enumerate}

Let us do so:

\begin{enumerate}
\item
We have $D=\diag(a,b,b)$. The Rescaling Lemma (Proposition \ref{rescaling}) allows us to choose $a=1$. 
Let $ \Omega=8\pi(\rho+p)\neq0$. The $QM$ equations are:

\begin{align*}
& QM.1 \sse \frac12 H^2= \Omega\, u^2 + \frac12 \,\Omega +8\pi p; \\
& QM.2 \sse H=-2\Omega\, u^0u; \\
& QM.3.ij \,\,(i\neq j) \quad \text{are already satisfied}; \\
& QM.3.ii \sse 
\begin{cases}
R_{11} = \frac12 = \Omega\, u^2 + 4\pi (\rho-p) \\
R_{22} = R_{33} = b-\frac12 = -\frac12 H^2 + 4\pi (\rho-p)
\end{cases}
\end{align*}
We then have
\[
QM.3.11+QM.3.22 \sse b=-\frac12 H^2 + \Omega u^2+8\pi (\rho-p).
\]

Inserting $QM.1$ in the last equation yields

\begin{align*}
& b=-\Omega \,u^2 -\frac12\,\Omega-8\pi p+ \Omega \,u^2 +8\pi (\rho-p) \\
& \sse b=-\frac12\Omega-8\pi p+8\pi (\rho+p)-16\pi p = \frac12\,\Omega -24\pi p\\ 
& \sse p = \frac1{24\pi}\left(\frac12\,\Omega-b\right).
\end{align*}

On the other hand,

\begin{align*}
QM.3.11 \sse \Omega\,u^2 & = \frac12 -4\pi(\rho-p) = \frac12 -4\pi(\rho+p)+8\pi p \\
& = \frac12 -\frac13\Omega - \frac13 b.
\end{align*}

Therefore,
\[
 u^2= \frac{3-2b}{6\Omega}-\frac13.
\]

Similarly,
\[
QM.1 \sse H^2 =\frac23(\Omega-2b)+1.
\]

Now
\begin{align*}
\hspace{3cm} & QM.2 \sse H=-2\,\Omega\, u^0u\\
&\se H^2= 4\,\Omega^2\left(u^0\right)^2u^2 \sse \frac1{4\,\Omega^2}H^2=u^4+u^2 \\
&\sse \frac1{4\,\Omega^2}\left(\frac23(\Omega-2b)+1\right)-\left(\frac{3-2b}{6\Omega}-\frac13\right)^2-\left(\frac{3-2b}{6\Omega}-\frac13\right)=0 \\
&\sse \left(-\frac{b}3+\frac14-\frac{(3-2b)^2}{36}\right)\frac1{\Omega^2}+\left( \frac16+\frac{6-4b}{18}+\frac{2b-3}6\right)\frac1\Omega+\frac29=0 \\
&\sse -b^2\frac{1}{\Omega^2}+b\frac{1}{\Omega}+2=0 \sse {1 \over \Omega}=\frac{-b \pm 3b}{-2b^2} \\
&\sse \Omega = \frac{b}2 \quad \text{or} \quad \Omega=-b.
\end{align*}

Let $\Omega=\frac{b}2$. We easily obtain
\begin{align*}
& H^2=1-b;\\
& u^2=\frac{1-b}{b};\\
& p=-\frac{b}{32\pi};\\
& \rho =-3p.
\end{align*}

To obtain the general solution , i.e., for $D=\diag(a,b,b)$,
we have to use the Rescaling Lemma. We have
\[
H^2(a,b,b) = a^2H^2\left(1,\frac{b}{a},\frac{b}{a}\right)=a^2\left(1-\frac{b}{a}\right)=a(a-b)
\]
yielding the condition $a\geq b$.

Similarly,
\[
u^2(a,b,b)=u^2\left(1,\frac{b}{a},\frac{b}{a}\right)=\frac{1-\frac{b}{a}}{\frac{b}{a}} = \frac{a-b}{b},
\]
yielding the condition  $b>0$. $QM.2$ requires only that $H$ and $u$ satisfy $H u \leq 0$. Finally,
\[
p(a,b,b)=a^2\left(-\frac{\frac{b}{a}}{32\pi}\right)=-\frac{ab}{32 \pi}.
\]
It can be shown that all these solutions of $QM$ are in fact Einstein's universe in
different frames (see section \ref{ID}).

If $\Omega = -b$, the procedure above yields the second family of solutions, corresponding 
to G\"odel's universe.

\item
We have $D=\diag(a,b,a-b)$. Let us set  $a=1$. The only changes 
with respect to the previous case occur in

\[
\hspace{3cm}
QM.3.ii \sse 
\begin{cases}
R_{11}=2b(1-b)=8\pi(\rho+p)u^2+4\pi(\rho-p) \\
R_{22}=R_{33}=0=- \frac12 H^2+4\pi(\rho-p) \sse H^2=8\pi(\rho-p)
\end{cases}
\]

From the last equation we obtain
\begin{align*}
\hspace{3cm}
QM.1 &\sse 4\pi(\rho-p)= 8\pi(\rho+p)u^2+4\pi(\rho+3p)\\
&\sse 8\pi(\rho+p)u^2=-16 \pi p \\
&\sse u^2= -\frac{2p}{\rho +p} \\
&\se (p\leq0 \quad \text{and} \quad \rho +p>0) \quad \text{or} \quad (p\geq0 \quad \text{and} \quad \rho+p<0).
\end{align*}

The second condition implies $\rho-p<-2p\leq0$, 
contradicting $H^2=8\pi(\rho -p)$. Since $\rho +p>0$, 
$H$ and $u$ must have opposite signs and

\begin{align*}
QM.2 &\sse \sqrt{8\pi(\rho-p)}=16\pi(\rho+p)\sqrt{1-\frac{2p}{\rho+p}}\sqrt{\frac{-2p}{\rho+p}}\\
&\sse \rho-p=64\pi p(p-\rho).
\end{align*}

But $\rho-p=0 \se H=0 \,{\buildrel QM.2 \over \Longrightarrow} u=0 \,{\buildrel QM.3.11 \over \Longrightarrow} R_{11}=0 \sse bc=0$, which is absurd.
We then have
\[
p=-\frac1{64 \pi}
\]
and
\[
QM.3.11 \sse \rho = \frac{32b(1-b)-5}{64\pi}.
\]

From the equations above we can then obtain the expression for $H^2$ and
$u^2$, in the special case $a=1$. To obtain the general solution
and the restrictions over $a$ and $b$, we proceed as in the previous case.
This third family of solutions can be shown to be the Ozsvath-Farnsworth-Kerr class I
family of solutions.
\end{enumerate}

\item \label{1b}
Let $u_1=0\,\, (\se H_1=0)$ and $\left(u^0 \right)^2 = -\frac{bc}{16\pi(\rho +p)}$.
We will prove that there are no solutions satisfying these hypotheses. 
We start by checking that
\[
\uu^2+1=-\frac{bc}{16\pi(\rho +p)} 
\sse 8\pi(\rho+p)\uu^2=-\frac12 bc-8\pi(\rho+p),
\]
and hence
\[
QM.1 \sse \HH^2=-bc-8\pi(\rho-p).
\]

On the other hand,
\[
QM.3.11\sse 8\pi(\rho-p)=R_{11}-\frac12 bc
\]
and so
\begin{align*}
QM.3.22+QM.3.33 &\sse R_{22}+R_{33} = 4\pi(\rho-p)-16\pi p\\
&\sse p= \frac{2R_{11}-4R_{22}-4R_{33}-bc}{64 \pi}.
\end{align*}

It is now immediate that
\[
\rho=\frac{10R_{11}-4R_{22}-4R_{33}-5bc}{64\pi}.
\]

On the other hand,
\begin{align*}
\hspace{3cm}
QM.2 \sse 
\begin{cases}
bH_2=-16\pi(\rho+p)u^0u_2 \\
cH_3=-16\pi(\rho+p)u^0u_3
\end{cases}
 &\se
\begin{cases}
b^2\left(H_2\right)^2=\left[16\pi(\rho+p)\right]^2\left(u^0\right)^2\left(u_2\right)^2 \\
c^2\left(H_3\right)^2=\left[16\pi(\rho+p)\right]^2\left(u^0\right)^2\left(u_3\right)^2
\end{cases} \\
 &\sse
\begin{cases}
\left(H_2\right)^2=-16\pi(\rho+p)\frac{c}{b}\left(u_2\right)^2 \\
\left(H_3\right)^2=-16\pi(\rho+p)\frac{b}{c}\left(u_3\right)^2
\end{cases}
\end{align*}

Using $\left(QM.2.2\right)^2$, we get

\begin{align*}
QM.3.22 &\sse R_{22}=\frac12 \left(H_2\right)^2-\frac12 \HH^2 +8\pi (\rho +p)\left(u_2\right)^2 + 4\pi(\rho-p)\\
&\sse \left(1-\frac{c}{b}\right)8\pi(\rho+p)\left(u_2\right)^2=R_{22}-R_{11}.
\end{align*}

It is easily checked that there are no solutions with $b=c$, and hence
\[
8\pi(\rho +p )\left(u_2\right)^2 =\frac{b}{b-c}(R_{22}-R_{11}).
\]

As a consequence of $\left(QM.2.2\right)^2$, we have
\[
\left(H_2\right)^2=\frac{2c}{c-b}(R_{22}-R_{11}).
\]

A similar procedure will give us
\begin{align*}
& 8\pi(\rho+p)\left(u_3\right)^2=\frac{c}{c-b}(R_{33}-R_{11});\\
& \left(H_3\right)^2=\frac{2b}{b-c}(R_{33}-R_{11}).
\end{align*}

From equation
\[
\left(H_2\right)^2 + \left(H_3\right)^2 = -bc-8\pi(\rho-p)
\]
we obtain the restriction
\[
-3a^2b+3a^2c+4ab^2-4ac^2-b^3-4b^2c+4bc^2+c^3=0.
\]

To simplify this last expression we use the Rescaling lemma to set $a=1$ and divide the
resulting polynomial equation by $b-c$, thus obtaining
\[
b^2+c^2+5bc-4b-4c+3=0.
\]

We have
\[
\left(u^0 \right)^2 = -\frac{bc}{16\pi(\rho +p)} \se bc(\rho + p)<0.
\]

The expression for $\left(H_3\right)^2$ implies that $\frac{b}{b-c}(R_{33}-R_{11})\geq 0$, and since
\[
8\pi(\rho +p )\left(u_2\right)^2 =\frac{b}{b-c}(R_{22}-R_{11}) \sse
8\pi bc (\rho +p )\left(u_2\right)^2 =\frac{b^2c}{b-c}(R_{22}-R_{11}),
\]
the restriction implied by the expression for $\left(u^0 \right)^2$ gives us
\[
\frac{c}{c-b}(R_{22}-R_{11})\geq 0.
\]

If we proceed in a similar fashion using the expressions for 
$\left(H_2\right)^2$ and $8\pi(\rho +p )\left(u_3\right)^2$, and then compute the 
components of the Ricci tensor in terms of $a$ and $b$, we will 
obtain the following restrictions:

\begin{enumerate}
\item
$b^2+c^2+5bc-4b-4c+3=0$;
\item
$c(c-b)(b-1)(b-c+1)\geq 0$;
\item
$b(b-c)(c-1)(c-b+1)\geq 0$;
\item
$bc(\rho + p)<0 \se 64\pi bc(\rho + p)<0 \se bc[3(b^2+c^2-3bc+4(b+c)-7]>0$.
\end{enumerate}

From (i) we obtain $b^2+c^2=-5bc+4b+4c-3$, which when used in (iv) yields
\begin{enumerate}
\item[(iv')]
$bc(-9bc+8b+8c-8)>0$.
\end{enumerate}

It is now easy to use a geometrical argument to determine the 
incompatibility of restrictions (ii), (iii) and (iv'): we just 
have to check that the regions determined in the $bc$-plane by these
restrictions do not intersect.
\end{enumerate}

\item
It is obvious that 
\[
ab=bc=ac \sse a=b=c.
\]
Symmetry allows us to consider  $u_2=u_3=0$, and thus we are back
to the very first case we analyzed.
\end{enumerate}
\end{proof}
%
%
%
\subsection{Solutions with $\GG=\OO$}

The next two results complete the classification of $QM$ class $A$
solutions with zero gravitational field.

\begin{Prop}
There are no class $A$ $QM$ solutions with zero gravitational field 
corresponding to Lie algebras with $\rank D=1$.
\end{Prop}

\begin{proof} Let $D=\diag(0,0,1)$. Symmetry allows us
to take $\HH=H_2X_2+H_3X_3 \buildrel QM.2 \over \Longrightarrow \uu=uX_3$.
Therefore, $QM.3.23 \sse 0=H_2H_3$.

If $H_3=0\buildrel QM.2 \over \Longrightarrow \uu=0$, we have
\[
QM.3.ii \sse \HH^2=-1.
\]

If $H_2=0$, we have
\[
QM.3.ii \sse 
\begin{cases}
H^2=8\pi(\rho-p)+1 \\
8\pi(\rho+p)u^2=\frac12-4\pi(\rho-p)
\end{cases}
\]
and 
\[
QM.1 \sse \rho=5p \se p\neq0.
\]

Using all this in  $QM2.2$ leads to $p=0$.
\end{proof}

\begin{Prop}
The only class $A$ solution of $QM$ with zero gravitational field 
corresponding to a Lie algebra with $\rank D=2$ is Minkowski spacetime.
\end{Prop}

\begin{proof} We can assume $a=0$. Thus $QM.2 \se u_1=0$, and therefore
\begin{align*}
\nabla_{\uu}\,\uu &= u_2u_3(\Gamma_{23}^1+\Gamma_{32}^1)X_1 \\
&= \frac12 u_2u_3 (C_{123}+C_{312}-C_{231}+C_{132}+C_{213}-C_{321})X_1\\
&= \frac12 u_2u_3(0-c+b+0+b-c)X_1 \\
&= (b-c) u_2u_3X_1.
\end{align*}

We then have
\[
\text{Motion equation} \sse 
\begin{cases}
(b-c)u_2u_3=u^0(u_2H_3-u_3H_2) \\
u_3H_1=0 \\
u_2H_1=0
\end{cases} 
\se H_1=0 \quad \text{or} \quad \uu=\OO.
\]
It can be easily seen that no solutions exist for $\uu=\OO$, and
that solutions featuring  $H_1=0$ and $u_2 u_3=0$ must
verify $b=c$, and hence are Minkowski spacetime (cf. Theorem \ref{flat}).

We are left with the case $H_1=0$ and $u_2 u_3 \neq0$. Using $QM.2$ we obtain
\[
\text{Motion equation} \sse QM.3.ij \,\,\,\, (i\neq j) \sse
\left(u^0\right)^2=-\frac{bc}{16\pi (\rho+p)} \,\,.
\]

The situation is now quite similar to the one in the demonstration 
of case \ref{1b} of Theorem \ref{rank3}. Using the same procedure we obtain
\[
u_2=\pm \frac{2\sqrt{3}}3 \sqrt{\frac{b^2}{-b^2-c^2+bc}} \se -b^2-c^2+bc>0
\]
and
\[
H_2= \pm \sqrt{-2bc} \se bc<0.
\]

But
\[
1 + \left(u_2\right)^2+\left(u_3\right)^2= -\frac{bc}{16\pi(\rho+p)} \sse b^2+c^2+5bc=0
\]
and, therefore,
\[
-b^2-c^2+bc>0 \sse b^2+c^2+5bc-6bc<0 \sse bc>0,
\]
yielding a contradiction.
\end{proof}
%
%
%
\subsection{Solutions with $\GG\neq\OO$}

For solutions with $\GG \neq 0$ we can assume $\GG=GX_1$ with $G \neq 0$, which implies $H_1 = u_1 = 0$ and $D=\diag(0,b,c)$. 
It is then easy to see that $QM1 + QM3.22 + QM3.33$ yields
\[
G^2 + 4\pi(\rho-5p) = 0.
\]
These solutions must of course include the two-parameter family given by
\begin{align*}
& H_2 = \sqrt{2} G \cos \theta; \\
& H_3 = \sqrt{2} G \sin \theta; \\
& u_2 = - \sin \theta; \\
& u_3 = \cos \theta; \\
& p = \rho = \frac{G^2}{16\pi}; \\
& b = c = 0,
\end{align*}
corresponding to the G\"odel universe. Apart from these, one can show that there 
exist further solutions, belonging to the category 2 of Ozsvath classification (see section \ref{ID}). 
Unfortunately, it is not possible to obtain simple expressions for these solutions.
%
%
%
\subsection{Identifying the solutions} \label{ID}

Recall that a solution of Einstein's field equations is said to be spacetime homogeneous if it 
admits a transitive action by isometries. This will happen if, for instance, the solution 
is a left-invariant metric on a (four-dimensional) Lie group.

The solutions we have been considering have in fact a Lie group structure, as 
$M=\bbR\times \Sigma$ and $\Sigma$ is a three-dimensional Lie group.

\begin{Prop}
A stationary spacetime $(M,g)$ corresponding to a solution of $QM$ for which the space 
manifold $(\Sigma,\gamma)$ is a Lie group with a left-invariant Riemannian metric and 
whose fields $\GG$ and $\HH$ are left-invariant is a Lie group with a left-invariant 
Lorentzian metric.
\end{Prop}

\begin{proof} 
One just has to check that
\[
\left\{ X_0 , X_i \right\}
\]
is a left-invariant orthonormal frame, where $\{ X_i \}$ are the vector fields in $M$ 
associated to a left-invariant orthonormal frame on the space manifold.
\end{proof}

Since all spacetime homogeneous perfect fluid solutions which are left-invariant 
Lorentzian metrics on a Lie group have been classified (see \cite{RS75}, \cite{SKMHH03}), 
we can use this classification to identify the solutions we have obtained. One must be 
careful to use frame-independent quantities when comparing solutions; in most cases it 
suffices to compare the equations of state.
%
%
%
\section*{Acknowledgments}
We would like to thank Filipe Mena and Jos\'e Mour\~ao for carefully reading an early version of this work.
%
%

\begin{thebibliography}{1}

\bibitem{B00}
J.~Bicak.
\newblock Selected solutions of einstein's field equations: their role in
  general relativity and astrophysics.
\newblock {\em Lect. Notes Phys.}, (540):1--126, 2000.

\bibitem{LN98}
D.~Lynden-Bell and M.~Nouri-Zonoz.
\newblock Classical monopoles: Newton, nut-space, gravomagnetic lensing and
  atomic spectra.
\newblock {\em Rev. Mod. Phys}, 70:427--446, 1998.

\bibitem{NZ97}
M.~Nouri-Zonoz.
\newblock Cylindrical analogue of nut space: spacetime of a line gravomagnetic
  monopole.
\newblock {\em Class. Quant. Grav.}, 14:3123--3129, 1997.

\bibitem{NZT01}
M~Nouri-Zonoz and A.~Tavanfar.
\newblock Plane symmetric analogue of nut space.
\newblock {\em Class. Quant. Grav.}, 18:4293--4302, 2001.

\bibitem{O02}
W.~Oliva.
\newblock {\em Geometric Mechanics}.
\newblock Springer, 2002.

\bibitem{RS75}
M.~Ryan and L.~Shepley.
\newblock {\em Homogeneous Relativistic Cosmologies}.
\newblock Princeton University Press, 1975.

\bibitem{SKMHH03}
H.~Stephani, D.~Kramer, M.~MacCallum, C.~Hoensalaers, and E.~Herlt.
\newblock {\em Exact Solutions of Einstein's Field Equations}.
\newblock Cambridge University Press, 2003.

\bibitem{W84}
R.~Wald.
\newblock {\em General Relativity}.
\newblock University of Chicago Press, 1984.

\bibitem{W83}
F.~Warner.
\newblock {\em Foundations of Differentiable Manifolds and Lie Groups}.
\newblock Springer, 1983.

\end{thebibliography}
%

\end{document}